\documentclass[preprint2]{emulateapj-rtx4}
\usepackage{graphicx,amsmath,url}
\usepackage{bm}
\usepackage{dcolumn}
\graphicspath{{./fig/}{./png/}}

\newcommand{\EQ}{\begin{equation}}
\newcommand{\EN}{\end{equation}}
\newcommand{\EQA}{\begin{eqnarray}}
\newcommand{\ENA}{\end{eqnarray}}

\newcommand{\Eq}[1]{Equation~(\ref{#1})}

\newcommand{\Sec}[1]{Section~\ref{#1}}

\newcommand{\Fig}[1]{Figure~\ref{#1}}

\newcommand{\Tab}[1]{Table~\ref{#1}}

{}
{}
{}

{}
{}
{}
{}
{}
{}
{}
{}
{}
{}
{}
{}
{}
{}
{}
{}
{}
{}
{}
{}

{}
{}
{}

{}

{}
{}

\newcommand{\vv}{\mbox{\boldmath $v$} {}}

\newcommand{\kk}{\bm{k}}

\newcommand{\xx}{\bm{x}}

\newcommand{\uu}{\mbox{\boldmath $u$} {}}

\newcommand{\parr}[2]{\frac{\partial #1}{\partial #2}}

\newcommand{\aaa}{\mbox{\boldmath $a$}{}}{}

\newcommand{\VV}{\bm{V}}

\newcolumntype{d}{D{.}{.}{-1}}

\begin{document}

\title{Turbulence Induced Collision Velocities and Rates between Different Sized Dust Grains}
\author{Alexander Hubbard$^{1,2}$}
\affil{
$^1$ Dept.~of Astronomy, American Museum of Natural History, 79th St at Central Park West, New York, NY
10024 \\
$^2$ Max Planck Institut f\"ur Astronomie, K\"onigstuhl 17, D-69117
Heidelberg, Germany \\
}
\email{alex.i.hubbard@gmail.com}

\begin{abstract}
We study the collision rates and velocities for point-particles of different sizes in
turbulent flows. We construct fits for the collision rates at specified velocities (effectively a collisional velocity
probability distribution) for particle stopping time ratios up to four; already by that point the collisional partners are very poorly correlated
and so the results should be robust for even larger stopping time ratios. Significantly, we find that while particles of
very different masses have approximately Maxwellian collisional statistics, as the mass ratio shrinks the distribution changes
dramatically.  At small stopping time ratios, the collisional partners are highly correlated and
we find a population of high number density (clustered), low relative-velocity particle pairs. Unlike in the case
of identical stopping time collisional partners, this low relative-velocity clustered population is collisional, but
the clustering is barely adequate to trigger bulk effects such as the streaming instability. 
We conclude our analysis by constructing a master fit to the collisional statistics as a function only of the stopping time
ratio. Together with our previous work for identical stopping time particle pairs,
this provides a recipe for including collisional velocity probability distributions in dust coagulation models for protoplanetary
disks. We also include
our recipe for determining particle collisional diagnostics from numerical simulations.
\end{abstract}

\keywords{turbulence -- planets and satellites: formation}

\section{Introduction}

Turbulent collisions between small dust grains is believed to lead to sticking and growth of the solids
in protoplanetary disks, a crucial early step in planet formation. However, as dust grains grow in size, they
respond more slowly to the gas motions.
This allows two neighboring dust grains to feel the effect of larger scale, higher velocity
gas motions differently, leading to higher collisional speeds.
Once the particles collide with large velocities, these collisions result in growth-neutral bouncing events
and, when even faster, fragmentation.
While the precise speed of turbulent motions in protoplanetary disks and the material strength of the dust is unknown,
the requirement that turbulence be adequately strong to explain observed accretion rates \citep{SS73, DiskLifetime}
imposes peak turbulent velocities noticeably above $10\,$m/s,
which itself is well above the dust fragmentation threshold of perhaps $1\,$m/s \citep{BW08, G10}.
It follows that there is a maximum
grain size beyond which equal-mass collisions will usually be too fast for the participants to survive; and so
such collisions lead to fragmentation rather than further dust grain growth \citep{DD05, Z10}.

Historically, most of the research into turbulence induced dust-dust collisions in astrophysics
 was spent analytically determining a single characteristic collisional velocity scale
for two dust grains of specified frictional stopping times with respect to the gas \citep{V80,M91,CH03,Y05,OC07},
but there has also been significant recent numerical work \citep{Carballido08,B10v, P11}.
 Given the large turbulent
velocities required to achieve significant and observed accretion flows in protoplanetary disks, this
has led to the conclusion that bouncing and fragmentation barriers to collisional dust grain growth are inevitable.
However, some recent work has instead focussed on the effect of including not merely a single
collisional velocity scale, but rather an entire collisional velocity probability distribution \citep{Okuzumi11}. Imposing Maxwellian
distributions around the previously estimated characteristic collisional velocities, they found
that small numbers of lucky grains who have multiple consecutive low-velocity encounters
could grow significantly past the bouncing and even fragmentation barriers \citep{Windmark12, Garaud12}.
This growth naturally peters out as large lucky grains are rare, and eventually unlucky collisions
destroy them. However, if these lucky grains become large enough compared to
the bulk population, which remains trapped by the bouncing barrier or ground up by the fragmentation barrier,
then they can grow even as the result of high velocity collisions with much smaller grains in a process
is referred to as mass transfer: they are large enough to survive the hits.
The grain growth process then is dominated by the sweep up of small grains
by a lucky few large dust grains, with equal mass collisions between the large grains being destructive but, fortunately, vanishingly
rare due to their small number density.

In a previous paper (\citealt{PaperI}, henceforth Paper I), we looked at the collisional velocity distribution for a monodisperse population
of dust grains (i.e.~a population with the same gas-dust drag parameters), and found a distribution that
was both lower in velocity than previously expected by a factor of approximately five, and very non-Maxwellian.  Indeed, in that case the
collisional velocity distribution had a pure exponentially decaying tail, and the particle pair velocity probability distribution 
appeared to diverge at low relative velocity.
In this paper we extend that analysis to
include polydisperse populations (collisions between grains of differing sizes), going from gas-dust
frictional stopping time ratios of
$17/16$ to $4$.  The stopping time scales with the grain radius and so the cube root of the mass, so our largest
stopping time ratio is equivalent to a mass ratio of $64$, large enough for mass transfer to occur.  It is also large enough
that we clearly approach an asymptotic limit as the two dust grains become uncorrelated.

Our goals are two-fold.  Firstly, we simply wish to understand any features and behavior of the collision velocity probability distribution
as there is no analytical treatment that can handle long time correlations of similar sized dust grains, and such correlations
were a dominant feature in the monodisperse case.  Secondly, we wish to develop a master fit for the collision velocity probability
distribution as a function only of the relative size of the dust grains.  Such a fit is of clear utility for any dust grain collision growth model
such as used in \cite{Windmark12} and \cite{Garaud12}.  We should note that our master fit only applies in the limit of the dust grains being well contained
within the turbulent cascade, as will be discussed below.

In Section 2, we discuss our turbulence model and how we extract the collisional diagnostics, going into greater detail on the
latter in the appendix. In Section 3, we perform initial analysis of our results, extracting clustering information as well
as the most common turbulent collisional speed. In Section 4 we examine the shapes of the collisional velocity probability
distributions and create fits to them which can be used in further modeling of dust coagulation. We conclude in Section 5.
In the Appendix we include a detailed procedure for extracting dust collisional diagnostics from numerical simulations.

\section{Methods}

As in Paper I, we evolve particles in a semi-analytic model for turbulence to extract collisional velocity probability distributions.
We do this for a range of stopping time ratios, constructing fits to the collisional velocity probability
distribution for each ratio, with a greater emphasis on identifying features in the distributions than
pure quality of fit. Finally, we construct
a fit to the fits to arrive at a final fit formula of the collisional velocity distribution
 with a small set of tunable parameters that can
easily to used in further particle coagulation studies such as \cite{Windmark12}.

\subsection{Particles}

We insert $10^6$ particles into the synthetic turbulence field. Writing $\uu$ for a dust particle's
velocity and $\vv$ for the gas velocity at the particle's position, the particle's equation of motion is
\EQ
\parr{\uu}{t}=-\frac{\uu-\vv}{\tau}, \label{dudt}
\EN
where $\tau$ is the particle's stopping time.  Note that $\tau$ is proportional to dust grain radius for
constant grain material density. 
All our simulations split the particles evenly between two species that differ only through their stopping times.
We refer to the larger stopping time as $\tau_1$ and the smaller as $\tau_2$.  Further, we define
\EQ
\epsilon \equiv \frac{\tau_1}{\tau_2}-1 \ge 0 \label{epsilon}
\EN
as the measure of the stopping time ratio.

The collisional properties of
a dust grain pair in incompressible turbulence are expected to scale with the stopping time of the largest grain because the shorter
stopping time grains are more tightly bound to the gas and particle collisions require deviations
from the incompressible gas velocity.
Further, dust is expected
to couple (for collisional purposes) most strongly with turbulence with turnover times $t_t=\tau$.
This is because if the turbulence has a turnover time $t_t \gg \tau$ the dust grains will be completely entrained
by the flow on a time scale $\tau$ and will barely deviate from the motion of the gas; while if the turbulence turnover time $t_t \ll \tau$ the dust grains will not have time
to respond to the gas motion.  Turbulent motions with timescales $t_t = \tau$ are therefore Goldilocks motions,
both varying rapidly enough that the particle motion can deviate from that of the gas, and varying slowly enough
that the particle motion is measurably affected.

We therefore denote $u_p$ and $k_p$
as the velocity scale and wavenumber of the turbulence with turnover time $1/k_p u_p=\tau_1= t_t(k_p)$. We denote
the associated velocity scale as $u_p$ (a dust velocity)
rather than $v_p$ (a gas velocity)
because this is the velocity scale we will use to normalize our dust collisional velocities.
When $k_p$ is not actually included in our turbulence model (which is
discrete in $k$), $u_p$ is determined by interpolating the Kolmogorov spectrum.
 
We consider modest $\epsilon \lesssim 3$ due to both practical considerations (numerical resources) and because
the larger $\epsilon$ case is expected to be straightforward.  That is because once the stopping time ratio
becomes large, the two populations of dust grains become uncorrelated, and, from the perspective of the
particles with stopping time $\tau_1 \gg \tau_2$, the grains with stopping time $\tau_2$ are completely bound to the gas motion
and so indistinguishable from the gas.
This means that the collisional behavior will have a well defined asymptotic behavior, which we find.

\begin{table}[b!]
\caption{Runs \label{Table:runs}}
\centerline{\begin{tabular}{lddddd}
\hline
 \multicolumn{1}{c}{Run} & \multicolumn{1}{c}{$\ \ \ \epsilon$} & \multicolumn{1}{c}{$\ \ \ \ \ \ \tau_1$}     
 & \multicolumn{1}{c}{$\ \tau_2$}& \multicolumn{1}{c}{$t_{ls}/\tau_1$} &\multicolumn{1}{c}{$\tau_2/t_{ss}$} \\
\hline
A      & 0.0625          & 2.01875 & 1.9        & 5.0             & 7.7 \\
B      & 0.125            & 1.96875 & 1.75      & 5.1              & 7.1 \\
C      & 0. 25              & 2              & 1.6        & 5                 & 6.5 \\
D      & 0. 375            & 2.3375   &  1.7       & 4.3              & 6.9 \\
E      & 0.5                 & 2.25        & 1.5         & 4.4              & 6.0 \\
F      & 0.9                 & 2.47        & 1.3         & 4.0             & 5.2 \\
G      & 1.25               & 2.7          & 1.2         & 3.7              & 4.8 \\
H      & 1.4                 & 2.88        & 1.2         & 3.5              & 4.8 \\
I      &  2                     & 3              & 1            & 3.3              & 4.0 \\
J      & 2.5                  & 3.5           & 1            & 2.9              & 4.0 \\
K      & 3                     & 4              & 1            & 2.5              & 4.0 \\
\hline 
\end{tabular}}
\end{table}

\subsection{Synthetic turbulence -- Motivation}

Since \cite{K41}, turbulence has been understood as a cascade of kinetic energy
from one lengths scale to a shorter one.  The largest length scale is the one associated
with the driving source of the turbulence, while the smallest length scale is determined
by the viscous dissipation of the turbulent energy into heat.  Between those two loosely defined
scales lies the inertial range.  In the inertial range the turbulence is scale free, independent of the details
of the forcing and of the dissipation: the only parameter is the energy cascade rate. The Kolmogorov power spectrum
of $v(k) \propto k^{-1/3}$ derived from the simplest dimensional analysis of the problem fits
experimental data well.
Significantly, in proto-planetary disks, the inertial range is believed to be large in that the ratio
of the time scale $t_{ls}$ associated with the largest scale motions to the time scale $t_{ss}$ associated
with the smallest scale motions is large: $t_{ls} \gg t_{ss}$.

From the hypothesis that particle-particle collisional properties are determined by turbulence
with $t_t \sim \tau_1$, and the consideration of modest $\epsilon$ it follows that there are
five regimes for turbulence induced particle-particle collisions, which can be describe by relating
$\tau$ to the bounding turbulent time scales, $t_{ls}$ and $t_{ss}$.  The regimes are:
\begin{itemize}
\item Particle stopping times much longer than the longest turbulent timescale, or $\tau \gg t_{ls}$: This regime may
be accessible to direct simulation of the turbulence using Navier-Stokes with current numerical resources, 
although the particles will move significant distances so the numerical domains will have to be large.
\item Particle stopping times comparable to the longest turbulent timescale, or $\tau \sim t_{ls}$: This regime should be accessible
to direct numerical simulation, although the applicability to protoplanetary disks may be complicated by
rapid particle drift through the disk, and the results may depend sensitively on the turbulent driving mechanism.
\item Particle stopping times comparable to the shortest turbulent time scale, or $\tau \sim t_{ss}$: This regime
is numerically accessible for hydrodynamical simulations, although the dissipation must be well captured.
\item Particle stopping times much shorter than the shortest turbulent time scale $\tau \ll t_{ss}$: This regime is numerically
accessible.
\item Particle stopping times much longer than the shortest turbulent time scale, but also much shorter than the
longest turbulent time scale, or $t_{ls} \gg \tau \gg t_{ss}$: This regime is where the bouncing/fragmentation barriers are believed to lie, 
but
is numerically inaccessible because such an inertial regime is beyond current numerical resources.
\end{itemize}

We are interested in the last regime, because full collisional velocity probability distributions are interesting precisely
when collisions at different speeds have different outcomes.  This regime is, however, not one that can currently be achieved
through direct numerical simulation. This is our motivation for using semi-analytical synthetic turbulence: the synthetic
turbulent cascade perfectly matches the Kolmogorov spectrum that fits experiment well while having a large
inertial range and being numerically feasible.

\subsection{Synthetic turbulence -- Model}

Our turbulence field uses the same approach as in Paper I: a synthetic turbulence field that matches the power spectrum
of the inertial range of Kolmogorov turbulence \citep{K41}.  In other words, while the absolute scale of the kinetic energy
is arbitrary, the relative energy in the motions at two given scales exactly matches that of perfect Kolmogorov turbulence,
which matches experiment well.

In this model, the turbulence is binned logarithmically in wave-vector space. A region of wave-vector space
with $|\kk| \in k_{\text{ref}} \pm \delta k$ is a spherical shell, so we refer to the bins as such, and the shell
$m$ is associated with $|\kk| = 2^m$.
Each shell $m$ is assigned a trio of wave-vectors $\kk_{mn}$ of the appropriate magnitude. The gas velocity
associated with each wave-vector
$\kk_{mn}$ is along a velocity unit vector $\hat{\vv}_{mn} \perp \kk_{mn}$, so that the final velocity field
is incompressible.

In Paper I we used multiple
methods for introducing time variation in the velocity field, including rotating the projection of the energy
in shell $m$ onto the trio of wave-vectors and varying the phase factor. In this paper, we will use the
rotation of the projection, along with a slow variation in phase.  The rotation of the projection occurs
as a random walk of length $2\pi$ over an eddy turnover time $t_m=1/v_m k_m$, and allows
the velocity at a specific position to rotate in space, avoiding steady velocity patterns.
The phase variation takes the form of a random walk of
the phase $\phi_{mn}$
of length $\pi$ over $3$ eddy turnover times. The final gas velocity field is
\begin{align}
\VV(\xx,t)  =\sum_{m,n=0,1}^{m,n=8,3} 
                  \sqrt{2} a_{mn}(t) v_m \hat{v}_{mn} \cos \left[\kk_{mn} \cdot \xx +\phi_{mn}(t)\right], \label{Vfield}
\end{align}
where $v_m$ is the velocity associated with the shell $m$
and  $\aaa_m(t)$ is a unit vector that rotates as a random walk over the turbulent time scale $t_m$.
We set $v_0=0.1$ in code units.
We refer the reader to Paper I for a more complete description and derivation.
We will discuss the need for the phase variation in \Sec{correlationlength}.

In Paper I, we found that the relative particle/gas motion in our model had reasonable statistics compared to
predictions (note that particle/gas
statistics are not yet well constrained as observed clustering immediately implies non-trivial
particle/gas correlations, \citealt{P11}); we also found that the particles were transported away from their
positions according to a random walk, appropriate for turbulent particle diffusion.
In this work, we also find in \Sec{PCV} that the relative
motion of particles with large stopping time ratios do approach naive expectations of uncorrelated
motion.

In all simulations in this paper, we include shells with $m$ ranging from $0$ to $8$ ($k_m=2^{m}$ extending
from $1$ through $256$).
The smallest wavenumber, $k_{m=0}=1$, corresponds to our box scale $L=2\pi$.
We set $t_{ls}=10$ and $t_{ss} =0.25$ in code units, 
where here $ls$ and $ss$ refer to the largest and smallest eddy scales included in the simulation, or $k=1, 256$.
Extending the Kolmogorov spectrum then, there is a $k$ value associated with any $t_{ls} > t_t > t_{ss}$ such
that $t_t=k^{-1} v(k)^{-1}$, although that $k$ value generally does not fall on one of the values associated with our
discrete shells.  In those cases we simply extrapolate the Kolmogorov spectrum.
The large value of $t_{ls}/t_{ss}=40$ allows us to fit a particle
pair with stopping time ratio of $\tau_1/\tau_2=4$ while maintaining at least $t_{ls}/\tau_1 \sim
\tau_2/ t_{ss} \sim 3$.  Our particle pairs are then contained within the effective turbulent cascade.
We list our stopping time choices in \Tab{Table:runs}.

\subsection{Dust snap-shots}
\label{diags}

Our particles are initialized with a random position and zero velocity. In Paper I, we found that it took a significant
time for the particle collisional diagnostics to achieve a steady state, in part due to the formation of dense clusters of highly
correlated particles. Following that result, we take snap-shots
every turbulent turnover time for the largest eddy, starting at the $20$th turnover. We then average over the snap-shots,
identifying small separation particle pairs and binning them in relative velocity (bin width $\delta u$).

For a detailed description of our procedure for extracting dust collisional diagnostics from this data, see the appendix. In brief,
we use the binned data to determine $N(R,u,j)$, which
is the number of particle pairs $(a,b)$ with separation $|\xx_a-\xx_b|\le R$, and relative velocity
$|\uu_a-\uu_b| \in u \pm \delta u/2$. The parameter
$j$ represents whether the particles have $\tau_a=\tau_b=\tau_1$ ($j=1$), $\tau_a = \tau_b = \tau_2$ ($j=2$) or $\tau_a \neq \tau_b$
($j=3$). Note that the largest $R$ we consider is $R_0$ with $R_0 k_{m=0}=0.02$.
Comparing with the minimum velocity limit $u_{min}$ from the appendix,
we note that the minimum relative velocity we can consider is
\EQ
\frac{u_{min}}{u_p} \sim 0.2 \frac{R_0}{\tau_p u_p} \sim 0.2 \frac{R_0}{L} \frac{L}{l_p} \gtrsim 7 \times 10^{-3}.
\EN
Our analysis remains are safely above this floor.

From $N$ we derive the fractional density enhancement over background
\EQ
\rho(R,u,j) = \left.{N(R,u,j)}\middle/\right.{\left[ \frac n2 \frac {4\pi R^3}{3} \frac {n}{L^3}\right]}/\delta u,
\label{rhoa}
\EN
where $n$ is the number of particles, $L^3$ is the box volume and we treat $\rho$ as a smooth function.
For both $N$ and $\rho$, a dropped $j$ implies that we are considering only collisions between particles of differing stopping times
($j=3$)
while a dropped $u$ implies summation over all velocity bins ($N$) or integration over velocity ($\rho$).
 A value of $\rho(R,j)=1$ is consistent
with randomly distributed particles while a value of $\rho(R,j)>1$ implies clustering on lengthscales of $R$ or less.

As discussed in Paper I, astrophysical particles are generally point particles in that their size is much smaller than that
of relevant gas dynamical length scales. Accordingly, we will attempt to determine the limit
\EQ
\rho(u,j) = \lim_{R \rightarrow 0} \rho(R,u,j).
\EN
The importance of this step can be understood through a consideration of the ``cold'' population found in Paper I.  This
population was large (it dominated the particle pair counts) but interacted with itself only with velocity $u_{\text{cold}} \propto R$.
The net result was that the cold population interacted at such low velocities that it provided no actual collisions despite
representing the bulk of the particle pairs.

 As long as both particles are
well trapped within the inertial range of the turbulence, then the problem becomes scale free because
the particles are too small to interact with the driving scale of the turbulence and too large to interact with the dissipative scale.
It follows that all
dust collisional parameters, once scaled to $u_p$, must be constructible from $\epsilon$, and knowledge of the
power law of the turbulent cascade.  This means that $\rho(u,j)$, for our fixed turbulent cascade, should be
independent of the precise values of $\tau_{1,2}$ so long as they are both smaller than $t_{ls}$ and $t_{ss}$
and $u$ is normalized to $u_p$.

\subsection{Scale free behavior}
\label{correlationlength}

As we have emphasized above, a crucial aspect of being well contained within the scale free
inertial range of the turbulence is that the particle-particle collision velocity probability distribution
must be scale free. Here we use the $j=1,2$ data to verify that our method produces
scale free results.

\begin{figure}[t!]\begin{center}
\includegraphics[width=\columnwidth]{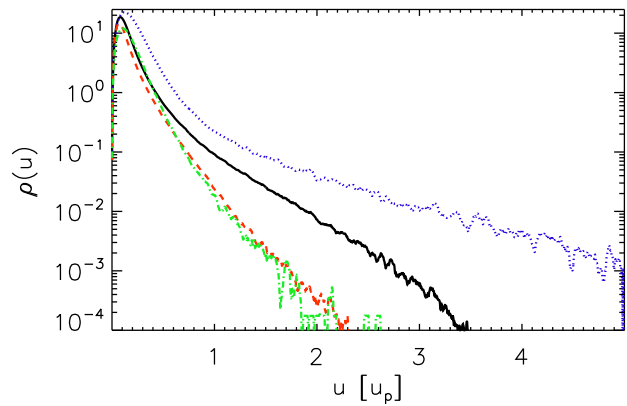}
\end{center}\caption{
Values of $\rho(u,j)$ from Run K with the time varying phase turned on (red/dashed, yellow/dash-dotted)
and off (black/solid, blue/dotted). The red/dashed and black/solid curves are for $R=R_0$, $j=1$ while
the green/dash-dotted and blue/dotted curves are for $R=R_0/2$, $j=2$.
Without the phase variation, we see an unphysical high $u$ tail which is not self-similar (different dependencies
on the appropriately scaled $u_p$). With the phase variation, this behavior is eliminated.
\label{phase} }
\end{figure}

Turbulent velocity fields with correlation length $\ell$ are expected to be uncorrelated on lengthscales
$l\gg \ell$. However, \Eq{Vfield} has infinite correlation lengths, which might be expected to set up
undesirable correlations between particles, which we do in fact find. In \Fig{phase}, which uses data from
both Run K and an equivalent run with the time-variation of the phase turned off, we show $\rho(R_0,u)$ for
$j=1$ (black/solid and red/dashed) and $\rho(R_0/2,u)$ for $j=2$ (blue/dotted and green/dash-dotted). The
black/solid and blue/dotted curves do not include the time-varying phase. The $x$-axis is plotted in units of $u_p$
for the particle pair, which is the gas velocity associated with turbulence with turnover time equal
to the particle stopping time, different for $j=1,2$.  This means that the velocity scaling factor for any given
particle pair is the one appropriate to that pair. Run K was chosen 
for illustrative purposes as it uses the largest and smallest values for $\tau_1$, $\tau_2$.

The size of $R_0$, chosen to guarantee a reasonable number of pair counts
at large relative velocity, is too large to derive any actual diagnostic
values from the curves plotted, but a clear difficulty emerges. When the phase variation is turned off,
a strong high velocity tail emerges, which is not scale free (i.e. does not scale with $u_p$, the turbulent
velocity scale that the particles interact with) as is shown
by the distinct slopes for the two particle populations $j=1,2$.
In practice, the tail appears to be controlled by the largest scale turbulence
included in the system. We attribute this problem to the infinite correlation length scale
of \Eq{Vfield} which allows particle motion to be correlated over long distances if the stopping time ratio
is close enough to $1$, while physical turbulence has well defined and finite correlation lengths.
When we consider polydisperse cases without time-varying phase,
the effect remains noticeable for $\epsilon \lesssim 0.25$, but larger stopping time ratios
break the correlations.

We break this correlation for all $\epsilon$ by adding the slow phase variation in \Eq{Vfield}.
While mutually approaching particles do sample the same infinite correlation length gas velocity
field, the nodes ($\vv_{mn}(\xx)=0$) that the two particles see will be in different locations as they cross positions
of the same $\kk \cdot \xx_p$ value at different time, with different phases $\phi_{mn}(t)$.
 In \Fig{phase} we see that
with the phase variation turned on, the particle pair distribution reverts to being scale free as required by
the turbulent cascade.

 This behavior
was not seen in Paper I even though there were setups where it should have occurred. This is due
to Paper I's low velocity bin cutoff and focus on small separations. At a given relative velocity $u=a u_p$, a particle
pair can only close a distance of order $\tau_1 u= a/k_p$.
At low relative velocities ($u<u_p$, the case in Paper I), particles can
only close separations smaller than the appropriate turbulent length scales, so the turbulent field we used
was appropriate.

\section{Existence of limits $R \rightarrow 0$}

Dust grains in protoplanetary disks, generally less than centimeters in size, are much smaller than the smallest
scale turbulent motions expected, which are larger than kilometer scales. Accordingly, actual collisions between such
dust grains cannot be resolved in a numerical simulation. As explained in \Sec{diags}, we therefore take the limit
$R \rightarrow 0$ for our diagnostic quantities.  For finite $\epsilon$ (unlike the $\epsilon=0$ case!),
this limit shows well defined and finite behavior for both  the most common relative velocity
and the total number of particle pairs.

\subsection{Peak collisional velocity}
\label{PCV}

\begin{figure}[t!]\begin{center}
\includegraphics[width=\columnwidth]{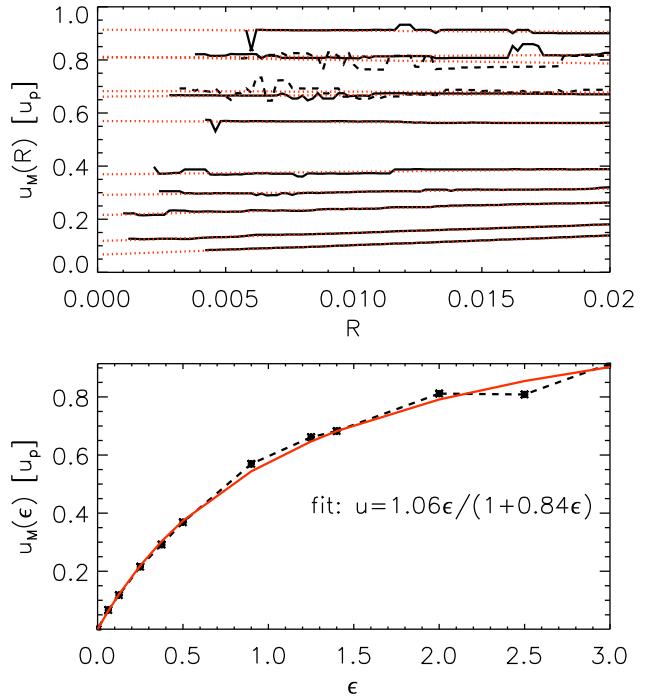}
\end{center}\caption{
Top Panel: $u_M(R)$, the velocity at which $\rho(R,u)$ has its maximum value when taken as a function of $u$ 
for a range of $R$ (black/solid except for
H and I which are black/dashed), along with 
their linear fits (red/dotted). There are finite, well defined limits of $u_M$ as $R \rightarrow 0$.
From bottom to top are runs A-K in order (H,I are dashed for clarity). The inner cut-off is where the pair counts become unreliable
for the fitting purposes.
Bottom Panel: Location of the peak (limit $R\rightarrow 0$) as a function of $\epsilon$.
\label{maxfit} }
\end{figure}

In Paper I, we found that the location $u_M$ of the peak of $N(R,u,j)$ in velocity space, for particles with the same stopping time,
is linear in $R$, with a $0$ intercept of $0$.  This implies the bulk of the particle population has a 
velocity distribution that is differentiable in space (i.e.~collisionless).
Allowing for different stopping times produces a different result, as shown
in \Fig{maxfit}, where the location of the peak remains finite even in the limit $R\rightarrow 0$.
For stopping time ratios close to $\epsilon \sim 0$, the location of the peak (at $R \rightarrow 0$)
 is approximately linear in $\epsilon$, while it levels
out for large $\tau_1/\tau_2$, as seen in the lower panel of \Fig{maxfit}.
This is expected as in that limit, much of the relative velocity comes from the
relative velocity between the larger dust grains and the gas (to which the smaller grains are tightly bound).

The fit in the bottom panel of \Fig{maxfit} implies a value of $u_M \simeq 1.3 u_p$ in the limit $\epsilon \rightarrow \infty$.
While we have not yet considered the full probability distribution of the collisional velocities, this is
of the same order as found analytically for this case; which is the limit for which theory should be most applicable
as the long-time correlations are the weakest. In particular, \cite{OC07} predict that the RMS-averaged collisional
velocity should reach $\sqrt{3} u_p$ in this limit, which is comparable to our limit for $u_M$.
We can also expect from \Fig{maxfit} that collisions will be significantly
slower than this value for $\epsilon \lesssim 1$. As particle mass scales with $\tau^3$, this corresponds to an already
noteworthy mass ratio $\sim 10$. One of the greatest difficulties in growing particles collisionally is the fragmentation barrier,
which is mitigated if the larger grain is larger enough to survive bombardment. While the collisional velocity we find
is lower than that previously anticipated for $\epsilon<1$, it is still high enough to lead to fragmentation for
 the largest scale turbulence expected in protoplanetary disks. Easing the collision velocity between near-equal mass
 grains should however lead to a larger production of ``lucky'' grains.

\subsection{Clustering}

Turbulence has long been known to expel heavy (higher density than the fluid) particles from high vorticity
regions through centrifugal forces and concentrate them in the high strain regions between vortices \citep{Maxey87,F94,C01,T09, P11}.
Indeed, particles with identical stopping time are strongly clustered in that the dust grain density perceived by a test grain
diverges to infinity as that grain considers ever smaller spheres around itself (\citealt{P11}, Paper I).
If such clustering persists for large ranges in the stopping time ratio, the streaming instability or gravitational collapse could be triggered 
\citep{GW73,J06,J07,Shi13}.
We should note here that
the vortices we consider are too small to feel any Coriolis forces that exist in a protoplanetary accretion disk.
\emph{Large} anti-cyclonic vortices in accretion disks are known to concentrate particles through orbital interactions\citep{BS95,J04},
but such effects require both rotation (which we do not include) and large enough length scales for orbital dynamics
to play a major role (which we do not consider).

\begin{figure}[t!]\begin{center}
\includegraphics[width=\columnwidth]{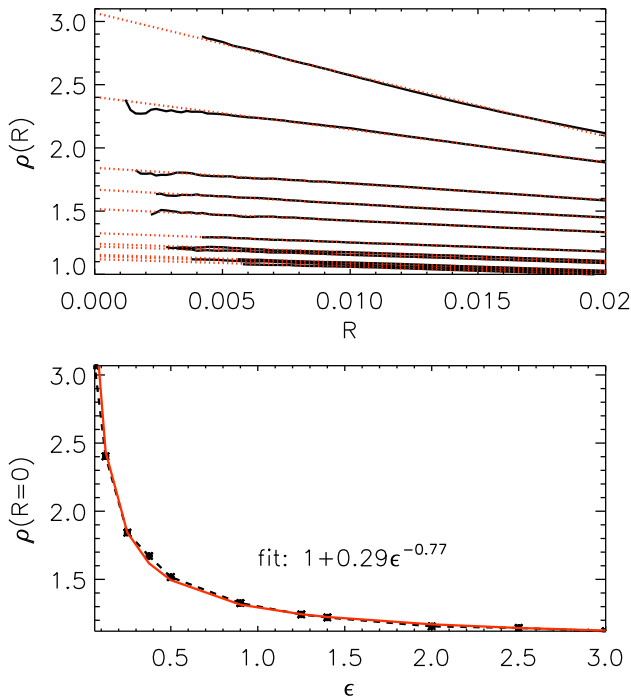}
\end{center}\caption{
Top Panel: Effective density enhancement $\rho(R)$ for a range of $R$ (black/solid), along with 
their linear fits (red/dashed). As in \Fig{maxfit} there are well defined limits as $R\rightarrow 0$, which
is not the case for $\epsilon=0$.
From top to bottom are runs A-K, in order. The inner cut-off is where the pair counts become unreliable
for the fitting purposes.
Bottom Panel: Limit $R\rightarrow 0$ of $\rho(R)$ as a function of $\epsilon$. The divergence is weaker than $\epsilon^{-0.77}$.
\label{cluster} }
\end{figure}

In \Fig{cluster} we show the clustering, i.e.\,$\rho(R)$, as a function of $R$ and $\epsilon$. In the case
where $\tau_1=\tau_2$, it has been found to behave as a power-law in $R$ with an exponent $\mu \sim -0.6$ \citep{P11}.
 Unlike that case, the density is not a power-law in $R$ when $\tau_1 \neq \tau_2$, instead having a well defined
 limit as $R \rightarrow 0$. Further, the clustering decreases as $\epsilon$ increases, as would be expected:
 the position of particles with significantly different stopping times are less correlated than those of particles
 with similar stopping times.

From \Fig{cluster} it is clear that the clustering diverges as $\epsilon \rightarrow 0$, as has been previously
found for $\epsilon=0$. However, because the clustering scales more weakly than
$\epsilon ^{-1}$ it is also clear, when combining that result with \Fig{maxfit}, that a net collision rate estimate
$u_M(\epsilon) \rho(\epsilon)$ does not diverge as $\epsilon \rightarrow 0$, which was also found in Paper I.
Nonetheless, the confirmation
of particle clustering even for different (but similar) stopping times continues to motivate the consideration
of physics relevant to  mostly-collisionless clusters of dust grains, which could otherwise have been dismissed
as a result specific to the monodisperse case.

It should be noted that the relatively weak power-law dependency on $\epsilon$ for the clustering
implies that the extreme levels of clustering predicted will not generate high
dust-to-gas mass ratio clumps.
For there to be significant direct back-reaction of the dust on the gas flow, the dust-to-gas mass ratio needs to be near unity.
If the volume averaged dust-to-gas ratio is the canonical $0.01,$ this implies a $100$--fold local increase in the dust
density.
To get a density enhancement over background of $100$ through turbulent clustering, 
we estimate a maximum $ \delta \epsilon$ range
of $0.29\, \delta \epsilon^{-0.77}=100$, or $\delta \epsilon = 5 \times 10^{-4}$, or a maximum mass ratio of
$1.5 \times 10^{-3}$. As long as the dust is not tightly concentrated at a specific grain size, the enhancement
to the dust-to-gas mass ratio from turbulent clustering will be well below $100$,
and we do not have any reason to expect turbulent concentration to directly alter the character of the turbulence,
meaning that our application of synthetic turbulence with no back-reaction possible is not immediately self-inconsistent.
The streaming instability becomes significant for dust-to-gas mass ratios of about $0.02$ \citep{J09}, double
the canonical disk-averaged value.  This would be expected for $\delta \epsilon \lesssim 0.08$, or a maximum
mass ratio of $1.25$.  Triggering the streaming instability through turbulent concentration may therefore be possible,
but it will require almost all the dust being found in grains of very nearly the same size.

\section{Velocity distribution fit}

It is now well known that the outcome of a collision between two dust grains can depend on the velocity
of the collision \citep{G10}.  This means that studies of collisional dust growth in protoplanetary disks
need to include estimates of the rates and collisional velocity probabilities, a step recently begun
by \cite{Windmark12} and \cite{Garaud12}.  However, those studies assumed Maxwellian collisional velocity
distributions, which is appropriate only for uncorrelated motion.  The clustering data of \Fig{cluster} immediately implies
that for modest $\epsilon$, a Maxwellian fit is inappropriate.

\begin{figure}[t!]\begin{center}
\includegraphics[width=\columnwidth]{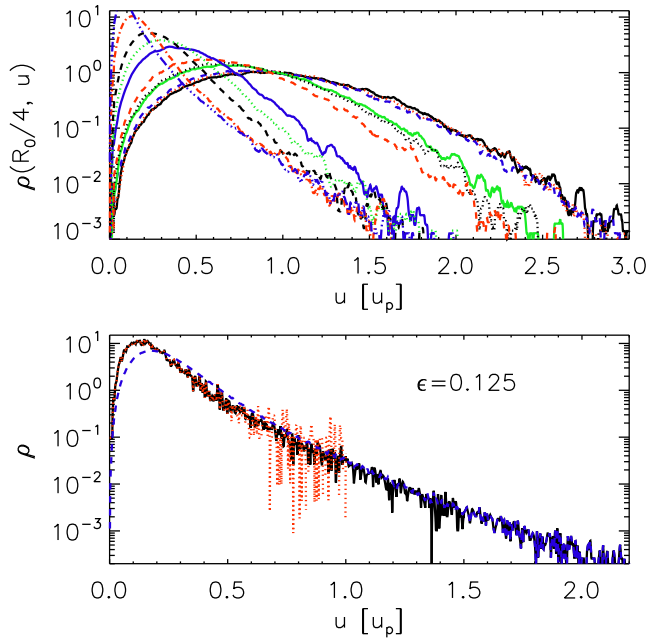}
\end{center}\caption{
Top panel: The particle pair normalized densities $\rho(R_0/4,u)$ as a function of $u_p$. From lowest
peak to highest we have runs A-K. There is a qualitative change as $\epsilon$ grows above $1$.
Bottom panel: $\overline{\rho}(u)$ (black/solid), $\rho(R_0/10,u)$ (red/dotted) and $\rho(R_0,u)$ (blue/dashed) for run B.
We can see that one needs to consider low values of $R$ when evaluating $\rho$ at low $u$, while larger
values of $R$ are both allowed, and required to sample an adequate number of particle-particle pairs when evaluating $\rho$ for large $u$.
\label{dist} }
\end{figure}

In \Fig{dist}, top panel, we show the normalized particle pair densities $\rho(R_0/4,u)$ as a function of $u_p$
for our runs. While $R_0 k_p$ is not a constant, and so we are not correctly normalizing our lengthscales to recover
perfectly
scale free behavior, we can nonetheless see a change in
behavior as the stopping time ratio $\tau_1/\tau_2$ increases, going from a sharply
peaked distribution around a relatively low velocity to a much broader distribution around a significantly higher
peak. We also have a qualitative change in the behavior of the high velocity tails, and the existence or non-existence
of a low velocity bulge, which occurs between $\epsilon=0.5$ and $\epsilon=0.9$.
Already by $\epsilon \gtrsim 2$ the high velocity tail appears to converge.

As alluded to in \Sec{correlationlength}, if a particle pair has relative velocity $u$, then they can close
a distance $\ell \sim \tau_1 u$. This means that the requirement of taking limits as $R\rightarrow 0$ is
less stringent for higher relative velocities. We illustrate this in the bottom panel of \Fig{dist}, where
we show results from Run B for $R=R_0/10$ and $R_0$, along with
\EQ
\overline{\rho}(u)=\begin{cases}
    \rho(R_0/10,u) & u/u_p<0.5 \\
    \rho(R_0/4,u) & 0.5<u/u_p<1 \\
    \rho(R_0/2,u) & 1<u/u_p<1.5 \\
    \rho(R_0,u) & 1.5<u/u_p. \end{cases}
\EN
This choice allows us to match
the need for small $R$ for low relative velocities while maximizing the particle pair count for
rare, high velocity pairs and avoiding sharp discontinuities. In what follows we always refer
to $\overline{\rho}$.

\subsection{Limiting cases}

Before attempting to fit the entire distribution for all our stopping time ratios with a universal
fit formula, we first consider subsections
of the distribution.

\subsubsection{Low relative velocity limit}

\begin{figure}[t!]\begin{center}
\includegraphics[width=\columnwidth]{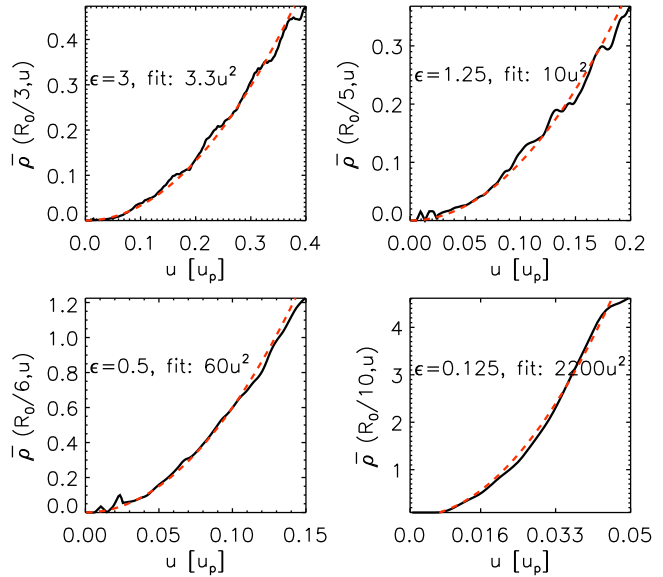}
\end{center}\caption{
Low $u$ limit of $\overline{\rho}(u)$ for four values of $\epsilon$.  All are approximately quadratic in $u$.
\label{lowu} }
\end{figure}

In the limit of low relative velocity $u$, we find that $\overline{\rho}(u) \sim u^2$ for both large and small $\epsilon$, 
as shown in \Fig{lowu}. This growth is faster for smaller $\epsilon$, which is reasonable as those
particle pairs are more tightly correlated, and will collide at lower velocities.
 Note that the increased clustering for small $\epsilon$ both allows us and
requires us to consider smaller values of $R$ than for large $\epsilon$. 

\subsubsection{Large $\epsilon$ -- Maxwellian}

\begin{figure}[t!]\begin{center}
\includegraphics[width=\columnwidth]{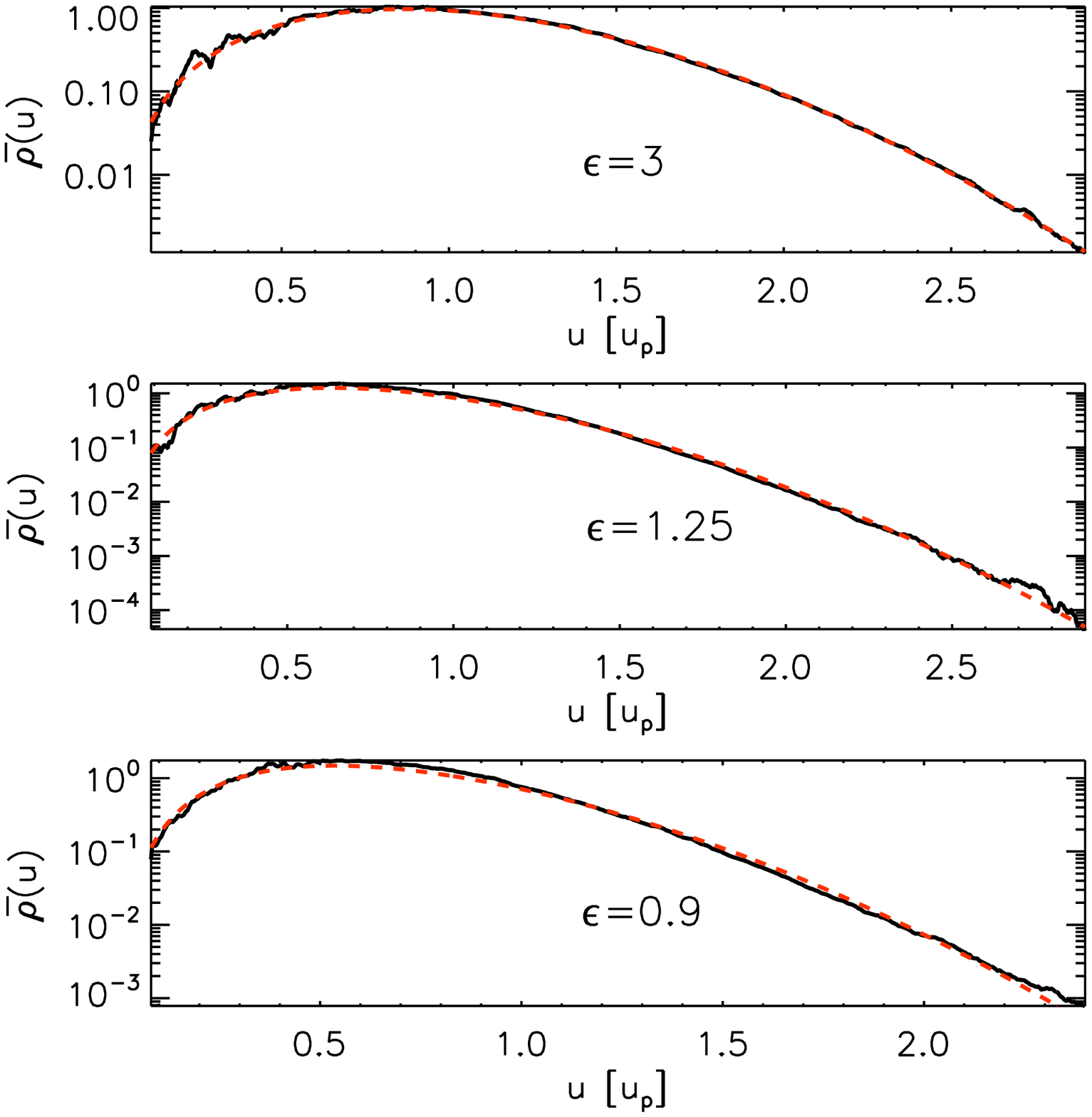}
\end{center}\caption{
Black/solid: $\overline{\rho}(u)$ for high $\epsilon$.  Red/dashed: fits using \Eq{highepsfitform}.
\label{higheps} }
\end{figure}

In the limit of large $\epsilon \gtrsim 1.25$, the distribution of particle pairs is well fit by an equation of the form
\EQ
\overline{\rho}(u) \sim A \frac{(B+1) C u^2}{B+u^2} e^{-C^2 u^2}, \label{highepsfitform}
\EN
as seen in \Fig{higheps}, where the fit function are the red dotted curves.
However for $\epsilon \lesssim 1.25$ this shape no longer fits the results, as the tail
flattens, approaching a linear exponential. \Eq{highepsfitform} is not quite a
Maxwellian probability distribution, but it is close: for $\epsilon \ge 1.25$ we have $B \sim 1$, so the fit is nearly
Maxwellian except for the high velocity tail. Note also that \Eq{Vfield} has a strict maximum which means that our
tail must eventually hit zero.

\subsubsection{Low $\epsilon$, high $u$ -- Exponential}

\begin{figure}[t!]\begin{center}
\includegraphics[width=\columnwidth]{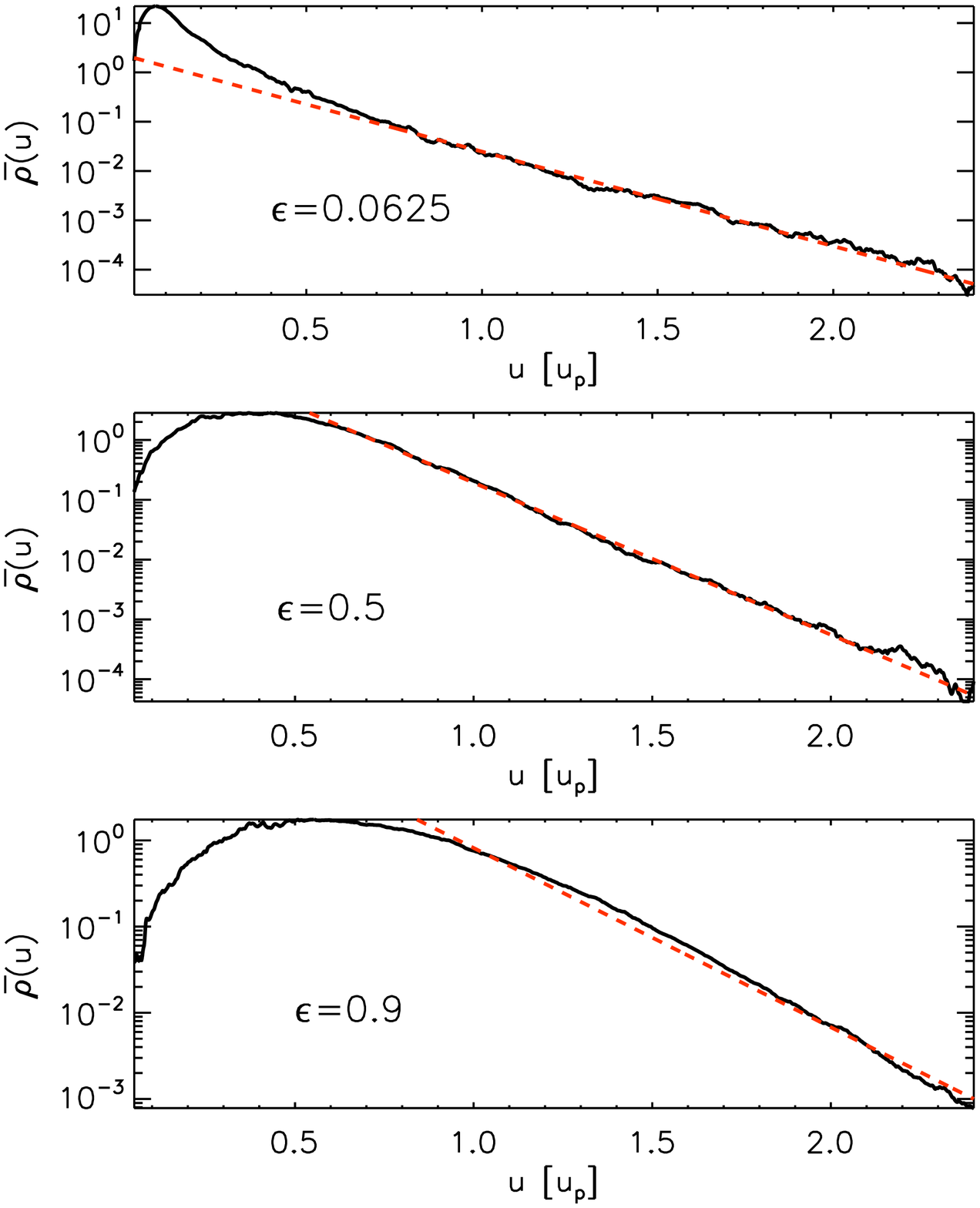}
\end{center}\caption{
Black/solid: $\overline{\rho}(u)$ for low $\epsilon$.  Red/dashed: fits considering only the high velocity tail, using \Eq{lowepsfitform}.
\label{loweps} }
\end{figure}

In the limit of high relative velocity $u$, we find
\EQ
\overline{\rho}(u) \sim A B e^{-Bu} \label{lowepsfitform}
\EN
is a good fit for low values of $\epsilon$ as shown in \Fig{loweps},
although our data range is too limited to constrain
a polynomial prefactor. We show fits for $\epsilon$ ranging from
$0.0625$ to $0.9$, and by the latter value the tail is clearly developing some curvature.

\subsubsection{Low $\epsilon$, low $u$ -- Clustering}

\begin{figure}[t!]\begin{center}
\includegraphics[width=\columnwidth]{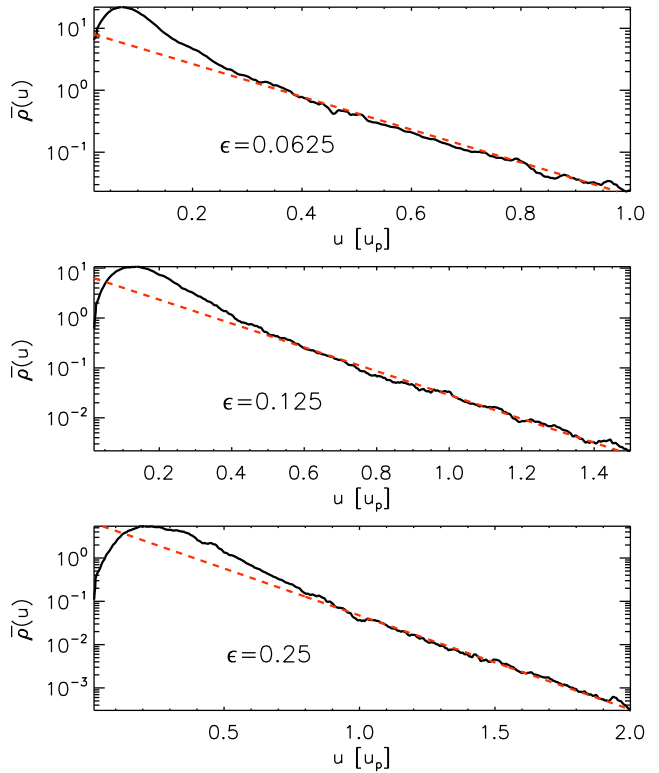}
\end{center}\caption{
Black/solid: $\overline{\rho}(u)$ for low $\epsilon$.  Red/dashed: fits considering only the high velocity tail, using \Eq{lowepsfitform}.
The difference between the two is evidence of the clustered population.
Note that we have zoomed in to lower $u$, and the zoom is stronger for lower $\epsilon$ due to the stronger, lower velocity
clustering.
\label{cluster2} }
\end{figure}

In \Fig{cluster2} we show low $u$ zooms of $\overline{\rho}$ for our smallest $\epsilon$ runs, along with exponential
fits for the higher $u$ tail. We can see that there is a distinct bump at low velocities that gets more pronounced
as $\epsilon$ gets smaller. In the $\epsilon \rightarrow 0$ limit,
this is the ``cold'' population of Paper I.  Unlike that case however, for $\epsilon \neq 0$ this bump in $\overline{\rho}$
is collisional as the peak collisional velocity $u_M$ is located in the bump and is finite (see \Fig{maxfit}, bottom panel).
We attribute this enhancement to the particle pair number density at low $u$ to turbulent clustering, and
therefore refer to the population as the ``clustered'' population.

\subsection{Global fit -- Form}

\begin{table}[b!]
\caption{Fitting formulas \label{Tab:ff}}
\centerline{\begin{tabular}{cc}
\hline
 Behavior & Fitting formula  \\
\hline
quasi-Maxwellian         & $f_M=a_M\frac{2c_M}{\sqrt{\pi}}\left(2b_Mc_M^2+1\right)\frac{u^2}{b_M+u^2} e^{-c_M^2u^2}$  \\
quasi-exponential  & $f_E=a_E c_E\left(\frac{b_Ec_E^2}{2}+1\right)\frac{u^2}{b_E+u^2}                    e^{-c_Eu}$           \\
clustered                & $f_C=a_C \frac{c_C^3 u^2}{2}                                                                                     e^{-c_C u}$          \\
\hline 
\end{tabular}}
\end{table}

\begin{figure}[t!]\begin{center}
\includegraphics[width=\columnwidth]{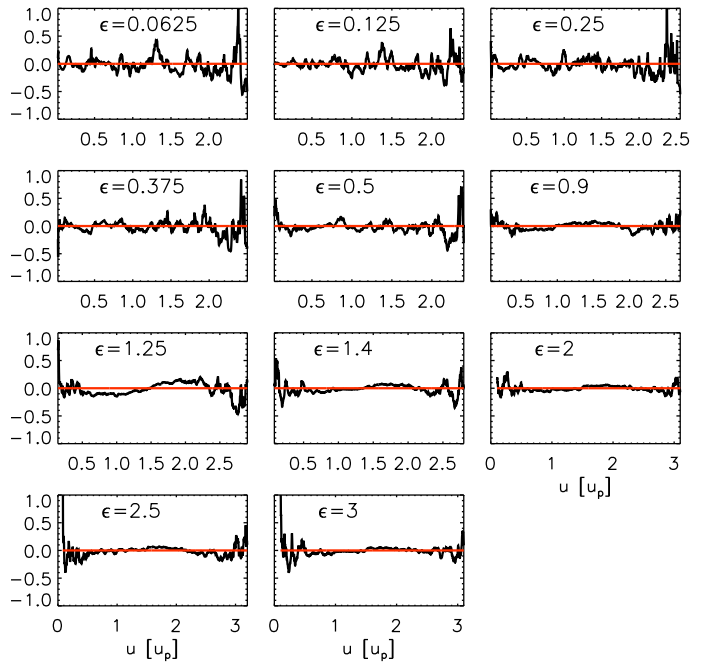}
\end{center}\caption{
Normalized errors for our fitting formula ($[f(u)-\overline{\rho}(u)]/\overline{\rho}(u)$) using formulas
 from \Tab{Tab:ff}.
The coefficients are determined for each run independently (listed in \Tab{Tab:vals}).
\label{errors2} }
\end{figure}

In this section we find a fit $f$ for the form of $\overline{\rho}$ so that
 $f(u) \simeq \overline{\rho}(u)$, that applies for all the $\epsilon$ we consider.
Note that in this section, we consider only the functional form of $f$, and allow it to depend on coefficients
that are determined for each value of $\epsilon$ individually.  We will fit the dependence of the coefficients on $\epsilon$
in the next section.

Combining the low $\epsilon$ results, the high $\epsilon$ results, and \Fig{dist}, we can see that there is a qualitative
change in the behavior around $\epsilon =1$, which corresponds to nearly a full order of magnitude
in the mass ratio of the particles. This is interesting as by that scale particle-particle collisions can
no longer be considered to occur between near-equals, and the larger grain has an increased chance
of surviving even fairly energetic collisions. This change in behavior can be understood through \Fig{maxfit}:
a particle pair with relative velocity $u$ samples a maximum separation of order $\tau_1 u$. Particle pairs with
$\epsilon<1$ see low velocity collisions that sample volumes well contained within a single eddy (or in our
synthetic turbulence, a single gas velocity wavelength), while
higher $\epsilon$ pairs can sample multiple eddies. This implies that analytical approaches to particle
collisional velocities can only be applied to large $\epsilon$ values: if particles begin their final approach
when well separated, they do not start strongly correlated and an ensemble averaging approach
can be used. If their final approach begins at a small separation however,
then the initial conditions are critical and contain long time correlations which cannot yet be analytically treated.

\begin{table*}[htbp]
\caption{Parameter fits \label{Tab:vals}}
\centerline{\begin{tabular}{ccccccccc}
\hline
$\epsilon$ & $a_M$ &     $b_M$ & $c_M$ & $a_E$ &    $b_E$ & $c_E$ & $a_C$ & $c_C$ \\
\hline
$0.0625$  &    $0.53$ & $0.0013$ & $2.9$ & $0.35$ & $0.0016$ & $4.2$ & $2.2$ & $0.066$  \\
   $0.125$ &     $2.4$ & $0.0088$ & $5.2$ & $0.049$ &    $13$ & $5.3$ & $0.59$ & $0.19$ \\
      $0.25$ &    $1.8$ &    $0.051$ & $3.0$ &   $0.11$ &   $9.0$ & $5.6$ & $0.41$ & $0.21$ \\
    $0.375$ &    $1.8$ &    $0.083$ & $2.4$ &   $0.68$ & $0.12$ & $4.9$ &      $0$ &       N/A \\
        $0.5$ &    $1.2$ &       $0.51$ & $2.2$ &   $0.51$ &   $4.0$ & $6.1$ &      $0$ &       N/A \\
        $0.9$ &    $1.6$ &       $0.43$ & $1.3$ &   $0.12$ &   $6.1$ & $4.2$ &      $0$ &       N/A \\
     $1.25$ &    $1.6$ &       $0.41$ & $1.2$ &          $0$ &      N/A &     N/A &      $0$ &       N/A \\
        $1.4$ &    $1.6$ &       $0.71$ & $1.2$ &         $0$ &       N/A &     N/A &     $0$ &       N/A \\
           $2$ &    $1.5$ &       $1.0$   & $1.0$ &          $0$ &      N/A &     N/A &      $0$ &       N/A \\
        $2.5$ &    $1.4$ &       $1.7$   & $1.0$ &         $0$ &       N/A &     N/A &     $0$ &       N/A \\
           $3$ &    $1.2$ &       $3.1$   & $1.0$ &          $0$ &      N/A &     N/A &      $0$ &       N/A \\
\hline
\end{tabular}}
\end{table*}

Our global fit model tries to capture the quasi-Maxwellian behavior at high $\epsilon$ in a term $f_M$, the quasi-exponential
high-$u$ tails at low $\epsilon$ in a term $f_E$ and the low velocity cluster for low $\epsilon$ in a term $f_C$, so our fit $f(u)$
for $\overline{\rho}(u)$ is made up of
\EQ
f(u)=f_M(u)+f_E(u)+f_C(u).
\EN
We will refer to the populations captured by the terms in order as quasi-Maxellian, quasi-exponential and clustered.
The forms for $f_M, f_E$ and $f_C$ are shown in \Tab{Tab:ff},
where the $a$ parameters are amplitudes (note that the quasi-Maxwellian and quasi-exponential tail components are imperfectly
normalized), the $b$ parameters are offsets and the $c$ parameters are inverse velocity scales. While the offset parameters
have proven important to the fits, and both the quasi-Maxwellian and quasi-exponential behavior is seen, eight
fitting parameters allow for too much freedom: fitting the parameters
for Runs A-K (using starting values from \Tab{Tab:parms}), we find values tabulated in \Tab{Tab:vals}, which shows some
trends, but is not generally directly useful (for example, the scatter in $a_E$ and $b_E$ is large). The fit quality is good however,
as seen in \Fig{errors2}, where we plot $[f(u)-\overline{\rho}(u)]/\overline{\rho}(u)$.

\subsection{Master fit}

The values for the coefficients in \Tab{Tab:vals} do not vary smoothly and monotonically with $\epsilon$, which
implies that our fit formula is not a perfect match for the data.  Nonetheless, we can construct a fit for the coefficients
as a function of $\epsilon$ which matches the data acceptably. Such a fit can then be adapted to generate an approximate
$\overline{\rho}(u)$ for any $\epsilon$, and so can be used in a particle growth model.

Requiring that for $\epsilon \le 0.5$, $c_C=2/u_M(\epsilon)$ which forces the peak of the clustered population to occur
for $u=u_M$, we can find fit formula in terms of $\epsilon$ alone, arriving at \Tab{Tab:parms} (certainly not unique), which identifies three regimes: a large $\epsilon$ regime where
the clustering term is insignificant, and two small $\epsilon$ regimes where it is important. Significantly, $a_E, b_E$ and $c_E$ do not vary
for small $\epsilon$: in this $\epsilon$ range the large $u$ tail is controlled by the exponential terms, and they have
converged as a function of stopping time ratio.

\begin{figure}[t!]\begin{center}
\includegraphics[width=\columnwidth]{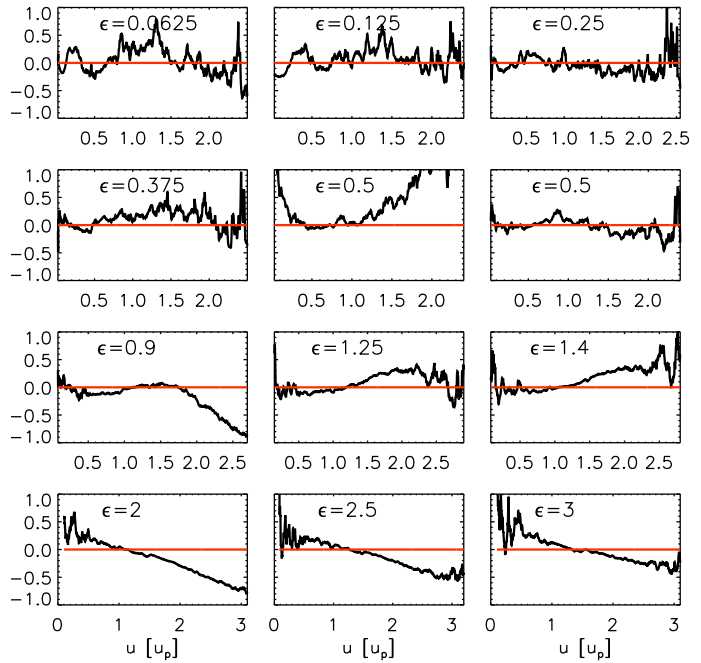}
\end{center}\caption{
Normalized errors for our fitting formula ($[f(u)-\overline{\rho}(u)]/\overline{\rho}(u)$) using 
formulas from \Tab{Tab:ff} with coefficients determined by the formulas in \Tab{Tab:parms}.  Note that
$\epsilon=0.5$ is shown with
both the lower $\epsilon$ fit (left) and upper fit (right). Compare with \Fig{errors2}, where the fit coefficients are determined for each
run independently.
\label{errors} }
\end{figure}

We show the error of this fit in \Fig{errors}, where we plot $[f(u)-\overline{\rho}(u)]/\overline{\rho}(u)$, which is, unsurprisingly, inferior to that in \Fig{errors2} because the latter has full freedom for the coefficients of $f$.
While for $\epsilon=1.25,\ 1.4,$ we underestimate $c_M$, it is clear that for larger $\epsilon$ our formula for $c_M$ does not approach
the unity found in \Tab{Tab:vals} rapidly enough. Less obviously, the formula for $f_M$ is imperfectly normalized.  We know from \Fig{cluster} that the integral
of $f(u)$ should be approximately $1$ in the large $\epsilon$ regime. For $\epsilon \lesssim 2.5$, we have $\int f_M(u) du < a_M$,
but for larger $\epsilon$ it rapidly approaches its asymptotic value of $a_M$. Those interested in such $\epsilon$ values should consider normalizing $f_M$ directly.

\begin{table*}[htbp]
\caption{Parameter function forms \label{Tab:parms}}
\centerline{\begin{tabular}{ccccccccc}
\hline
 $\epsilon$ range                &                                    $a_M$ &                $b_M$ &                                                       $c_M$
                                                &                                    $a_E$ &                 $b_E$ &                                                        $c_E$
                                                &                                    $a_C$ &                                                                                        $c_C$ \\
\hline
$\epsilon \ge 0.5$               &                                       $1.6$ &  $0.45 \epsilon$ &                    $1+\frac{1}{4 \epsilon^2}$ 
                                                & $0.04 \frac{1}{\epsilon}^3$ &                      $7$  &                                                            $6$ 
                                                &                                          $0$ &                                                                                              N/A \\ \\
$0.25 \le \epsilon < 0.5$   &                                       $1.6$ &     $0.3\epsilon$ &                         $\frac{6}{1+4\epsilon}$ 
                                                &                                       $0.5$ &               $0.33$  &                                                         $4.6$ 
                                                &         $\frac{0.08}{\epsilon}$ &                                $\frac{2+1.68 \epsilon}{1.06 \epsilon}$ \\ \\
$\epsilon \le 0.25$               &                                       $1.6$ & $4.8 \epsilon^3$&                         $\frac{6}{1+4\epsilon}$ 
                                                &                                       $0.5$ &                 $0.33$  &                                                        $4.6$ 
                                                &          $\frac{0.08}{\epsilon}$ &                                $\frac{2+1.68 \epsilon}{1.06 \epsilon}$ \\
\hline 
\end{tabular}}
\end{table*}

\section{Conclusions}

We have extended the work of Paper I to find turbulence induced collision rates and velocities between particles
with different stopping times (i.e.~of different sizes). Broadly, we find that the collisional
velocity probability distribution is approximately Maxwellian for large stopping time ratios, but deviates strongly from that limit for particles
with stopping time ratios less than $2$ (which corresponds to mass ratios below $10$). We also find, similarly to Paper I, that our collisional
velocities are smaller than previously estimated  \citep{OC07} for particles with modest stopping time ratios, although the previous estimates
are not bad when the particles are significantly different in size.

Further, we have managed to construct a usable estimate for the
collision rates and velocities between dust grains of different sizes in turbulent flows which captures their qualitative behavior for
arbitrary stopping time ratio (see Paper I for the case of monodisperse particles).

We interpret our results by noting that particle pairs with stopping time ratios close to unity (so particles of similar sizes)
collide at low speeds. This means
that they started their ``final approach'' while close together.  The change in behavior between quasi-Maxwellian and
non-Maxwellian collisional velocity probability distributions
occurs approximately when particle pairs start their final approach one eddy size apart. This
is the difference between particles whose positions and velocities have been correlated for a significant time, and
particles which are nearly uncorrelated. Under that interpretation, it is hardly surprising that the collisional velocity
probability distributions are different, nor that the large stopping time ratio collisions are quasi-Maxwellian.

We find a counterpart to the cold, collisionless population of Paper I (collisions between particles of identical stopping times),
in our clustered population.  In the case of non-unity stopping time ratios however, this population is collisional: the finite difference
in stopping times forbids perfect correlation between the dust grains. We also find that while particles of similar stopping times
are strongly concentrated by the turbulence, this clustering is unlikely to generate local dust grain density enhancements high
enough above background to lead to streaming instabilities, although it is within the realm of the possible.

While the different collisional velocity probability distributions for different particle size ratios is not surprising, it is of particular interest
for dust coagulation models. This is because the only effect that changing the velocity scale of collisions would have is to shift
the final dust grain size, and shifting the final grain size enough that grains grow large enough to trigger gravitational instabilities or similar is implausible.
Instead we want to enable more lucky dust grains to sneak through the fragmentation barrier by a fortuitous sequence of low velocity
collisions before growing large enough to sweep up small grains in high velocity encounters.

\acknowledgements

This work was supported by  a fellowship from the Alexander von Humboldt Foundation.
Further support was provided by
NSF Cyberenabled Discovery Initiative grant AST08-35734, and a Kalbfleisch Fellowship from the American Museum of Natural History.
We thank Jeffrey Oishi and Mordecai-Mark Mac Low for providing helpful comments.
\appendix

\section{Measuring dust concentrations and collisions}

In protoplanetary disks, dust grains that interact hydrodynamically with turbulence
are far smaller than the smallest turbulent eddies, unlike the situation in many
other systems including rain droplets in the Earth's atmosphere \citep{S03,X08}.  This makes it impossible
to simultaneously resolve individual particle-particle collisions along with the turbulence that
stirs them. This complicates the interpretation of numerical simulations designed to study
the rates and velocities of turbulence induced particle-particle collisions. Rather than actually
tracking individual collisions, we must instead take periodic snap-shots of the entire
system, identify interesting particle-particle pairs and convert their number, separation and
relative velocities into global collision rates and collisional velocities. On the other hand, the same
small size of the dust grains along with their high material density compared to the gas
means that we can neglect the volume displaced by the grains,
which is a significant simplification.

To convert snap-shots of the system to collisional diagnostics, we must make the fundamental assumption
that for any relative velocity $u$, there is a size scale $\ell_d$,
simultaneously much smaller
than the smallest turbulence length $\ell_{\nu}$ and much larger than the dust grain radii $a$, where
in all spheres of radius $R<\ell_d$ constructed around test dust grains,
the number density of dust grains with velocity $u$ with respect to the
 test grains is independent of $R$. This assumption will be broken if new physics emerges when dust
grains approach each other closely (such as mutual repulsion of negatively charged dust grains; \citealt{Okuzumi09}).  If dust grains are large
enough to induce wakes in the gas that affect particle-particle collisions, our assumption will also break down.

\subsection{Parameters and considerations}

The particles (moving with velocity $\uu$) and the gas (with velocity $\vv$) interact through the drag term
\EQ
\frac{\partial \uu}{\partial t} = -\frac {\uu-\vv}{\tau_p},
\EN
where $\tau_p$ is the friction stopping time. Assume a turbulent cascade where
the velocity of the turbulence at length-scale $\ell$ is $\vv_{\ell}$ and the turbulent
correlation time (turnover time) at that scale is $t_{\ell}=\ell/\vv_{\ell}$. While it has not yet been rigorously determined,
the strong expectation is that the turbulence which is most strongly coupled to particle-particle
relative motion (as opposed to bulk turbulent transport) is the turbulence at the scale $\ell_p$
where $t_{\ell}=\tau_p$ \citep{V80}. In the case of pairs of particles with different stopping times it is generally
believed that the largest stopping time defines the most-coupled turbulence.

\subsection{Sampling method}

As in Paper I and this work, take snap-shots of the particles with temporal separation $\delta t$, and bin particle pairs in space
and relative velocity (velocity bin width $\delta u$).
 From the binned data calculate $N(R,u,t)$, which is the number of particle pairs $p_1$ and $p_2$ with $|\xx_1-\xx_2| < R$,
$|\uu_1-\uu_2| \in u \pm \delta u /2$, in the snap-shot taken at time t. A dropped $u$ implies summation over
velocity bins
while a dropped $t$ implies averaging over snap-shots.

Note that
\EQ
N(R,u,t)/\left[\frac 12 n_p \times \left(\frac{4\pi}{3} R^3 \right)\right]
\EN
is the effective number density of particles at relative velocity $u\pm \delta u/2$ in spheres of radius $R$ as seen by particles
 in the snapshot $t$. The volume averaged number density of particles $\rho_0$ in a numerical simulation is however arbitrary
 (especially if they do not back-react on the gas), so it, along with the also arbitrary size of the velocity bin, should be scaled out:
 \EQ
\rho(R,u,t) \equiv N(R,u,t)/\left[\frac 12 n_p \times \left(\frac{4\pi}{3} R^3 \right) \rho_0 \right]/\delta u
\EN
(see also \Eq{rhoa}, this paper) measures the density of targets with relative velocity $u$ within distance $R$ compared to the background number density.
 Similarly to $N$, for $\rho$ a dropped $u$ implies integration over velocity (giving the ratio of perceived number density to background) while a dropped $t$ implies a time average over snapshots.
 A dropped $R$ will be discussed in further detail below, but implies a limit $R\rightarrow 0$.

\subsection{Sampling rate}

If the relative velocity of a particle pair is $u$ and $2R/  \delta t > u$, then that pair
can appear in multiple consecutive snap-shots. This can result in over/undersampling particles in regions of particle over/underdensities.
If we have a large number of snap-shots taken over a total time interval $t \gg R/u$,
our fundamental assumption implies that the oversampling of regions of over/underdensities will cancel.
Nonetheless, it is safer to choose a sampling rate with $\delta t > 2 R_{max}/u_{min}$ where $R_{max}$ is the
largest particle-particle separation we consider, and $v_{min}$ is the smallest relative velocity we consider.

More importantly, turbulence generates significant structure in the particle distribution, which can take multiple
particle stopping times or turbulent turn-overs to saturate, and leads to large swings in $N(t)$, as seen in Paper I. As such, it is crucial to sample across multiple particle
stopping times and multiple turbulent turnovers of the largest eddies which significantly interact with the dust grains.

\subsection{Sampling radii}

If $R/u > \tau_p$, then the drag force will significantly affect the relative motion of particle pairs with a separation $R$
before they could collide. This implies that for any collisional sphere radius $R$ considered, there is a minimum
velocity $u \gtrsim R/\tau_p$ for which one can extract collisional information. Given that particle relative velocities
 are expected to couple
most strongly to turbulence with turnover time $t_{\ell_p} \simeq \tau_p$, this immediately implies that a collisional sphere
of the same size scale of that turbulence, $\ell_p=v_{\ell_p} t_{\ell_p}$, will not be able to resolve collisional velocities
of $v_{\ell_p}$ or less.
This is a strong constraint: to calculate collision rates and collision velocities (especially modest velocities)
for dust grains, we need to resolve particle
separations much smaller than the length scale associated with the turbulence expected to maximally couple to
the particle-particle relative motions. See the discussion of the ``cold'' population in Paper I for a particularly interesting consequence for particles with equal stopping times.

\section{Method}

Taking the above considerations into account, one arrives at the procedure for extracting collisional diagnostics
from the snap-shots.

\begin{itemize}
\item Examine the time series of $N(R,u,t)$ to verify that one is sampling a particle population whose collisional
parameters have achieved a steady state, and that the sampling window is long enough to average over
long-time fluctuations.

\item Take the limit $\rho(u_{\star}) = \lim_{R \rightarrow 0} \rho(R,u_{\star})$ for all values of $u$ where this is possible, taking into account
the requirement that $u_{\star} \gg R_{min}/\tau_p$. Where this is possible, the assumption is that one has achieved $R<\ell_d$.
Where it is not possible, either one has $R>\ell_d$ or the length scale $\ell_d$ does not exist. In the latter case, it is 
not clear how collisional diagnostics could be designed without resolving individual particle-particle collisions.
\end{itemize}

\subsection{Concentrations and rates}

\begin{itemize}

\item $\rho_0 \rho(R)$ is the number density of particles within spheres of radius $R$ seen by test particles. If $\rho(R)>1$ particles
are clustered, while if $\rho(R)<1$ they are segregated at a lengthscale $R$.

\item $\rho_0 \rho(u_{\star})$ is the effective number density of particles with relative velocity $v_{\star}$ that interact with
test grains. Note both that it has units of density times inverse velocity and that
it is not the term to use for collision rates!

\item $\sigma \rho_0 u_{\star} \rho(u_{\star})$ is the contribution of the collisional velocity $u_{\star}$ to the collision \emph{rate} for a single particle, where $\sigma$ is the particle-particle
collisional cross-section.
Note the factor of $u_{\star}$, which
accounts for the fact that particles moving slowly with respect to one another will collide only on long time scales.

\item $\sigma \rho_0 \int_u du\ u_{\star} \rho(u_{\star})$ is the total collision rate for a single particle
\item $\left\{\left[\int_u du\ u_{\star}^3 \rho(u_{\star})\right]/\left[\int_u du\ u_{\star} \rho(u_{\star})\right]\right\}^{1/2}$ is the 
root-mean-squared particle-particle collisional velocity.

\end{itemize}

%
%
%


\end{document}